\documentclass[amsmath,amssymb,superscriptaddress,preprintnumbers,twocolumn,aps,nofootinbib]{revtex4-2}

\usepackage{graphicx}
\usepackage{dcolumn}
\usepackage{easyReview}
\usepackage[unicode]{hyperref}
\usepackage[capitalise]{cleveref}
\usepackage[utf8]{inputenc}
\usepackage[greek,english]{babel}

\usepackage{physics}
\usepackage{comment}
\usepackage{enumitem}
\usepackage{todonotes}
\usepackage[normalem]{ulem}
\usepackage{subcaption}
\usepackage{tikz}
\usepackage[compat=1.1.0]{tikz-feynman}
\usepackage{ragged2e}
\usepackage{fontawesome5}

\usepackage{ctable} 
\usepackage{multirow}
\usepackage{bm}

\newcommand{\lett}{\textit{letter}}
\newcommand{\met}{\textit{MET}}
    
\usepackage{eso-pic}

\begin{document}

\usetikzlibrary{decorations.pathreplacing}

\preprint{UMD-PP-023-04}
\preprint{MI-HET-817}

\title{A new purpose for the $W$-boson mass measurement:\\
searching for New Physics in lepton+$\met$}

\author{Kaustubh Agashe}
\email{kagashe@umd.edu}
\affiliation{Maryland Center for Fundamental Physics, Department of Physics, University of Maryland, College Park, MD 20742, USA
}

\author{Sagar Airen}
\email{sairen@umd.edu}
\affiliation{Maryland Center for Fundamental Physics, Department of Physics, University of Maryland, College Park, MD 20742, USA
}

\author{Roberto Franceschini}
\email{roberto.franceschini@uniroma3.it}
\affiliation{Universit\`{a} degli Studi  and INFN Roma Tre,\\Via della Vasca Navale 84, I-00146, Rome}

\author{Doojin Kim}
\email{doojin.kim@tamu.edu}
\affiliation{Mitchell Institute for Fundamental Physics and Astronomy, Department of Physics and Astronomy, Texas A\&M University, College Station, TX 77843, USA}

\author{Ashutosh V. Kotwal}
\email{ashutosh.kotwal@duke.edu}
\affiliation{Department of Physics, Duke University, 
Durham, NC 27708}

\author{Lorenzo Ricci}
\email{lricci@umd.edu}
\affiliation{Maryland Center for Fundamental Physics, Department of Physics, University of Maryland, College Park, MD 20742, USA
}

\author{Deepak Sathyan}
\email{dsathyan@umd.edu}
\affiliation{Maryland Center for Fundamental Physics, Department of Physics, University of Maryland, College Park, MD 20742, USA
}

\date{\today}

\begin{abstract}

We show that the $m_W$ measurement is a {\em direct} probe of New Physics (NP)
contributing to $\ell+\met$, independently from {\em indirect} tests via the electroweak fit. 
Such NP modifies the kinematic distributions used to extract $m_W$, necessitating a simultaneous fit to $m_W$ and NP. 
This effect can in principle bias the $m_W$ measurement, but only to a limited extent for our considered models. 
Given that, we demonstrate that the agreement at high-precision with SM-predicted shapes results in bounds competitive to, if not exceeding, existing ones for two examples: anomalous $W$ decay involving a $L_{\mu} - L_{\tau}$ gauge boson and $\tilde{\nu}_{l} \tilde{l}$ production in the MSSM.

\end{abstract}

\maketitle

\section{Introduction}

The mass of the $W$ boson plays a crucial role in our understanding of nature. The discrepancy between the recent and most precise measurement by CDF \cite{CDF:2022hxs} and the SM prediction might already be a hint of new physics (NP) beyond the Standard Model (BSM). Theoretical explanations commonly invoke new contributions to the electroweak (EW) fit \cite{deBlas:2022hdk} in order to shift the value of the SM prediction (see for instance \cite{Strumia:2022qkt,Asadi:2022xiy}) and explain the anomaly.
Yet, the more recent re-measurement by ATLAS \cite{ATLAS-CONF-2023-004,ATLAS:2017rzl} adds to the puzzle, confirming the SM-predicted value and the previous measurements by LHCb, D$\emptyset$ and LEP \cite{LHCb:2021bjt,D0:2012kms, ALEPH:2006cdc}. Whether in the future the CDF anomaly will be confirmed cannot be foreseen. The only fact that we have today is the striking precision of $10^{-4}$ of these measurements and of the corresponding theory SM predictions. This precision might even improve in the near future due to an ongoing intense experimental \cite{ATLAS-CONF-2023-004,Manca:2800948} and theoretical effort (see e.g. Refs.~\cite{Campbell:2023lcy,Camarda:2023dqn,Autieri:2023xme,Rottoli:2023xdc,Isaacson:2022rts,Chen:2022lpw,Chen:2022cgv} for recent works).

The $m_W$ experimental value is extracted from the simultaneous fit of different measured kinematic distributions (see below) in leptonic decays of singly-produced $W$-bosons to the SM predictions. 
Both ATLAS and CDF find perfect agreement with their best-fit SM distributions.

We show in this \lett~that the data used for the $m_{W}$ measurement can simultaneously be a powerful direct probe for any NP that contributes to the same final state. The key observation is that NP produces kinematic distributions that are sufficiently different with respect to those in the SM.
Hence, the same analysis can be used for the extraction of both $m_W$ and NP parameters. The correct procedure thus requires a global fit, which might in principle shift the measurement of $m_W$, with NP providing new nuisance parameters. 

\begin{figure}[t!]
    \center
    \includegraphics[width=0.45\textwidth]{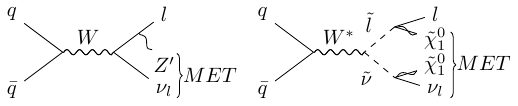}
 \captionsetup{justification=Justified, singlelinecheck=off}
\caption{NP contributions to the $W$-boson mass sample in the $\ell+{\rm \met}$ channel. Left: invisibly-decaying $L_{\mu}-L_{\tau}$ $Z'$-boson. Right: slepton-sneutrino production in the MSSM.\label{Fig:FeynDiag}}
\end{figure}

This paradigm is general, having already been attempted in \cite{Han:2012fw,Czakon:2014fka, Eifert:2014kea, Cohen:2019ycc,Franceschini:2015kdm,Bagnaschi:2023wip,Franceschini:2023nlp} for the top quark, in the context of NP copiously produced via strong interactions. Fainter signals of NP charged only under the electroweak interaction are more challenging. Yet we will show how the extraordinary precision of the $m_W$ measurement can put competitive bounds on motivated new physics scenarios, and in some cases to {\it exceed} present bounds, e.g. those for long-sought SUSY sleptons. This strategy is in addition to the classic test based on EW fit of the SM to which we are accustomed since LEP \cite{LEPEWWG}.
In this \lett, we focus solely on the $m_W$ measurement. We classify the possible NP that can contaminate the measured sample and quantify the sensitivity to two concrete, well-known BSM scenarios (see~\cref{Fig:FeynDiag}). 

\begin{figure*}
    \centering
    \begin{minipage}{.5\textwidth}
        \centering
        \includegraphics[width=0.9\linewidth]{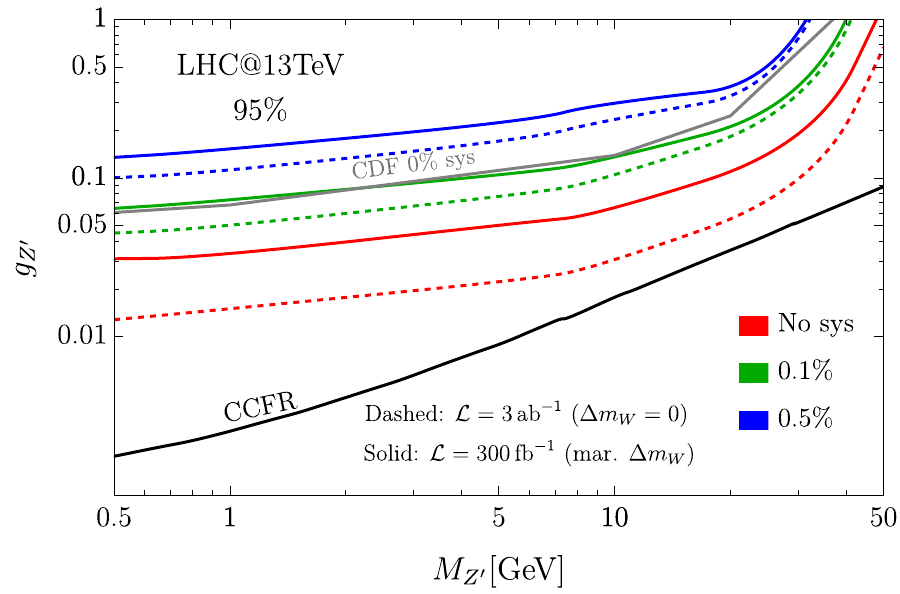}
        \subcaption[0.5\textwidth]{$pp \rightarrow \mu \,\nu_{\mu}\, Z'$}
        \label{subfig:zprime_money}
    \end{minipage}%
    \begin{minipage}{.5\textwidth}
        \centering
        \includegraphics[width=0.9\linewidth]{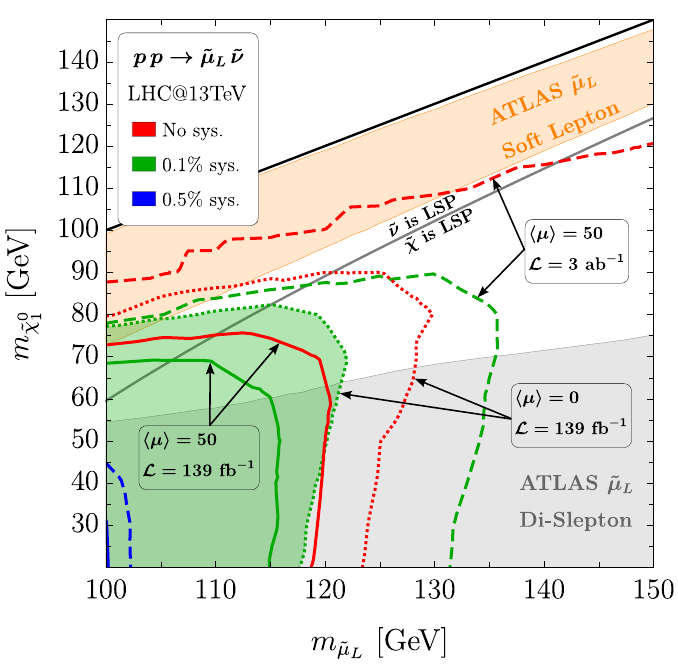}
        \subcaption[0.5\textwidth]{$pp \rightarrow \tilde{\mu}\, \tilde{\nu}_{\mu}$}
        \label{subfig:susy_money}
    \end{minipage}
    \captionsetup{justification=Justified, singlelinecheck=off}
    \caption{LHC 95\% CL projected sensitivity to (a) $L_{\mu}-L_{\tau}$ and (b) MSSM slepton-sneutrino production. All the lines include detector simulations. Pileup ($\langle \mu \rangle = 50$), simulated through the dedicated \textsc{Delphes} ATLAS card, is included unless indicated otherwise. In the SUSY projections, we include the no pileup ($\langle \mu \rangle = 0$) lines only for the competitive run-2 projections. Present bounds are obtained from \cite{Altmannshofer:2014pba} and \cite{ATLAS:2022hbt} respectively for the left and right figure.}
    \label{fig:money_plots}
\end{figure*}

\section{Invisible New Physics behind the semi-invisible W-Boson}\label{Sec:NPtoWMass}

The $W$-boson mass measurement is special. The remarkable precision, reached by hadron colliders, relies only on the partially visible leptonic decays. The masses of other heavy SM bosons are instead extracted from fully visible and clean final states (e.g., $h\to \gamma\gamma$, $Z\to\ell^+\ell^-$), hence resonance reconstruction is possible in a narrow region. For hadronic $W$-boson decays, resonance reconstruction is plagued by the challenges of QCD observables. The semi-invisible final state of leptonic $W$-decays, namely $\ell+\met$, is cleaner, but it presents a good hideout for invisible NP.

Given that the $W$-boson decay cannot be fully reconstructed, the measurement of the $m_W$ is a result of the fit to the lepton $p_T^{\ell}$ and the transverse mass $m_T$ distributions.\footnote{CDF also fits the missing transverse momentum $p_T^{\text{miss}}$ distribution.} 
Hence, any BSM that contributes to the same final state, modifying these kinematic distributions, can affect the $m_W$ measurement. Such NP can be classified in three possibilities: 
\begin{enumerate}[label=(\Alph*)]
    \item anomalous $W$-boson decay,
    \item anomalous $W$-boson production,
    \item $\ell+\met$ not from an on-shell $W$-boson, $\ell= (e,\mu)$.
\end{enumerate}
The first (second) possibility includes all BSM models that modify the $W$-boson decay (production), yet resulting in $\ell+\met$. Option (C) collects all BSM models that can produce an $\ell+\met$ final state, without the involvement of any on-shell $W$-boson. This category includes the production of new particles, decaying into $\ell+\met$, and new interactions among quark/gluons and leptons.\footnote{Examples of this are dim-6 quark-lepton four fermion operators that mediate $qq\rightarrow \ell\, \nu_\ell$ processes. The latter are usually very well constrained by high-energy measurements \cite{Farina:2016rws,Torre:2020aiz,Panico:2021vav}.}

Here we explore two simple, yet relevant, case studies that cover options (A) and (C). In Sec.~\ref{Sec:lmultau}, we focus on anomalous $W$-boson decay in the invisibly-decaying $L_{\mu}-L_{\tau}$ gauge boson scenario (Fig.~\ref{Fig:FeynDiag} left). This represents a proof-of-principle of our idea, highlighting the relevant points with rather simple phenomenology. Nevertheless, we find that the $m_W$ measurement represents a competitive probe for this model (see Fig.~\ref{subfig:zprime_money}). In Sec.~\ref{Sec:SUSY} we focus on category (C), using $\tilde{\nu}\tilde{\ell}$  production in SUSY as an example. This production mechanism is not currently investigated at the LHC.  Remarkably, our results in Fig.~\ref{subfig:susy_money} show that the $m_W$ measurement can cover an unexplored parameter space of slepton searches.

In a follow-up paper \cite{Agashe:toappear}, we will study additional examples of category (A) and an illustration of category (B): a $Z^{ \prime }$-boson gauging baryon number (see \cite{Dobrescu:2021vak} and references therein). Overall, our two papers thus represent a {\em comprehensive} study of probing NP giving $\ell+\met$ using $m_W$ analysis. Ref. \cite{Bandyopadhyay:2022bgx} studied a specific example of category (B) only. Moreover, in the following, we describe a more general approach than Ref.~\cite{Bandyopadhyay:2022bgx} for the associated analyses.

\section{\boldmath A proof-of-principle: \texorpdfstring{$L_{\mu}-L_{\tau}$}{L\textmu - L\texttau} gauge boson}\label{Sec:lmultau}

The first model that we consider is the $L_{\mu}-L_{\tau}$ $Z'$ \cite{He:1991qd}:
\begin{align}
    \mathcal{L}_{\text{int}} = g_{Z'} Z'_{\rho} J^{\rho}_{\mu - \tau}+g_{D}Z'_{\rho} J^{\rho}_{D}\,,
\end{align}
where $g_{Z'}$ and $g_D$ are the couplings of $Z'$-boson to SM and dark-sector states, respectively. 
The $U(1)_{L_{\mu} - L_{\tau}}$ current reads
\begin{eqnarray}
   J^{\rho}_{\mu - \tau} &=& (\bar{\nu}_\mu \gamma^\rho \nu_\mu+\bar{\mu} \gamma^{\rho} \mu -\bar{\nu}_\tau \gamma^\rho \nu_\tau -\bar{\tau} \gamma^{\rho} \tau).
\end{eqnarray}
The term $Z'_{\rho} J^{\rho}_{D}$ describes the interaction of the $Z'$-boson with some invisible, unspecified dark-sector states. The key assumptions, that $g_{D}\gg g_{Z'}$ and the dark sector contains states sufficiently lighter than $m_{Z'}$, guarantee that the $Z'$-boson decays predominantly invisibly.
 
This model has been extensively studied as a possible portal to dark matter or as an extension to SM. The 2-dimensional parameter space $(g_{Z'},m_{Z'})$ is tested by a variety of searches, from K-/B-factories, $g-2$, to neutrino beam-dump experiments \cite{Altmannshofer:2014pba,  Krnjaic:2019rsv}.\footnote{Additional constraints arise when $m_{Z'}$ is of Stuckenberg origin \cite{Ekhterachian:2021rkx}.} In this model belonging to category (A), the $W$-boson has a 3-body decay into $\mu \,  \nu_{\mu} \, Z'$ (Fig.~\ref{Fig:FeynDiag} left), modifying the kinematic distributions of $\ell + \met$ final state.\footnote{Additional signal events come from $\tau\rightarrow Z' \mu \, \nu_{\mu} \, \nu_{\tau}$. For simplicity we don't include them in our analysis.}

\begin{figure}
    \centering
\includegraphics[width=0.45\textwidth]{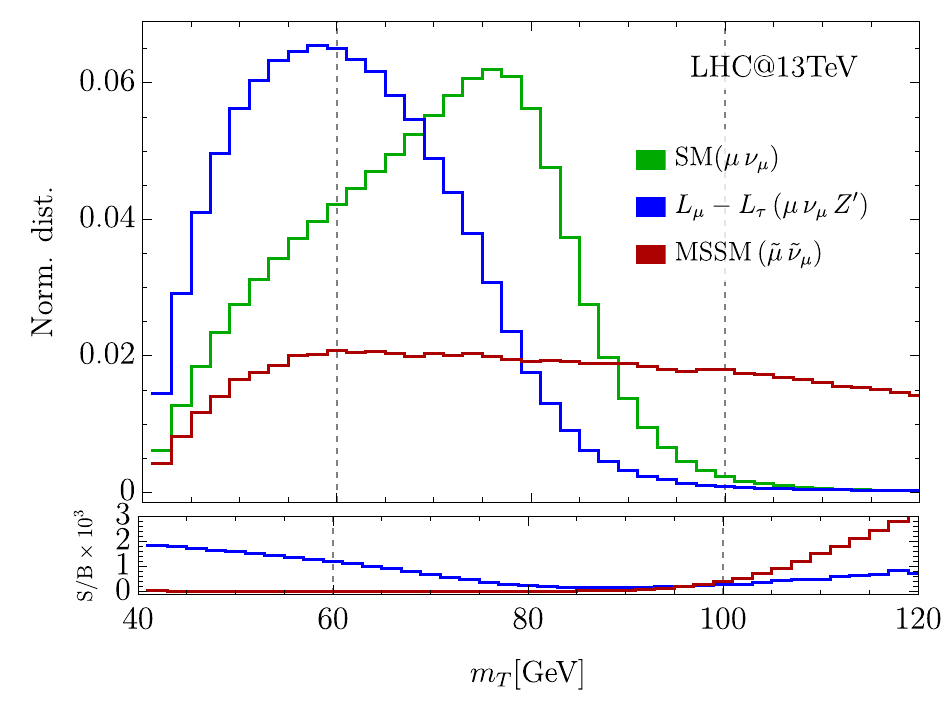}
\captionsetup{justification=Justified, singlelinecheck=off}
    \caption{Normalized transverse mass distributions for $\mu + \met$ at the LHC. Blue line: $m_{Z'} = 10\,\text{GeV},g_{Z'} = 0.12$).  Red line: $m_{\tilde{\mu}} =  115 \, \text{GeV}$, $m_{\tilde{\nu}} = 83\, \text{GeV}$, $m_{\tilde{\chi}^{0}_{1}} = 70 \,\text{GeV}$. 
    The dashed lines in the lower panel are obtained from selected $Z$ events. The dashed gray lines indicate the ATLAS fitting range.}
    \label{fig:Distrib}
\end{figure}

We obtain the kinematic distributions through a Monte Carlo (MC) simulation via \textsc{MadGraph5\_aMC@NLO}v3.42 \cite{Alwall:2014hca} + \textsc{PYTHIA8.212} \cite{Bierlich:2022pfr} + \textsc{Delphes}v3.4 \cite{deFavereau:2013fsa} (ATLAS card). We employed \textsc{LHAPDF} \cite{Buckley_2015}, PDF  ID:244800 \cite{Ball_2013}.
The 3-body decay (versus 2-body) softens the $p_T$ and $m_T$ distributions, as seen in Fig.~\ref{fig:Distrib} for a benchmark value of $(m_{Z'},g_{Z'})=(10~{\rm GeV},0.12).$\footnote{NP also modifies $W$-boson total decay width. This effect is expected to be negligible given the projected bound on the NP parameters. Therefore we fix the width to its SM value. The effect of the width on the $m_W$ determination within the SM is only a few MeV. \cite{Isaacson:2022rts,ATLAS-CONF-2023-004}.}

As shown in Fig.~\ref{fig:Distrib}, for $g_{Z'}\sim \mathcal{O}(0.1)$, the expected $S/B$ ratio is $\mathcal{O}(10^{-3})$. Sensitivity to these effects strongly relies on the various sources of uncertainties, which is exactly the main target for the experimental collaborations that reached percent \cite{CDF:2022hxs} and even sub-percent uncertainties \cite{ATLAS-CONF-2023-004,ATLAS:2017rzl}, aimed at measuring $m_W$. Also backgrounds are extensively studied and they are only a few\% in the region of interest. In this \lett~we will not attempt a complete study of the various sources of uncertainties in the presence of NP. We just comment on the possible effect of our NP hypothesis on the sample of $Z\rightarrow \ell \ell$ events which are heavily used for detector calibration~\cite{ATLAS:2017rzl,CDF:2022hxs} and for tuning the boson production model on data~\cite{Isaacson:2022rts}. 
Thus a contamination of NP in the $Z\to\ell\ell$ sample might affect the calibration of the MCs, ``calibrating away'' signs of NP~\cite{Hammou:2023heg}.
However, by isolating pure $Z$-boson events with appropriate kinematic cuts, such as those imposed by ATLAS \cite{ATLAS:2017rzl}: $80<m_{\ell\ell}/\text{GeV}<100$, the possible contamination of NP in the calibration sample is limited to $\mathcal{O}(10^{-4})$, still for $g_{Z'}\sim \mathcal{O}(0.1)$. 

We estimate the sensitivity and the impact of our NP hypothesis on the $m_W$ measurement through a binned $\chi^2$ analysis for the $p_T^\ell$ and $m_T$ distributions. Our analysis is aligned as much as possible with the ATLAS measurement \cite{ATLAS-CONF-2023-004,ATLAS:2017rzl}, only slightly extending the fit range aiming at maximal sensitivity (see Tab.~\ref{tab:CUTS}). 
We then construct the following $\chi^2$:
\begin{align}
    \chi^2(\Delta_{m_W} , \Delta_{\text{NP}}) = \sum_{i =1}^{N_{\text{Bins}}} \frac{\left(N_{ev}^i(\Delta_{m_W} , \Delta_{\text{NP}})-\overline{N}_{ev}^i  \right)^2}{\sigma^2_{stat}+\sigma^2_{sys}}\,,
    \label{Eq:ChiSqu}
\end{align}
where $N_{ev}^i (\Delta_{m_W}, \Delta_{\text{NP}})$ is the expected number of events in the the bin $i$ as function of $m_W$ ($\Delta_{m_W} = m_W - \overline{m}_W$) and the NP parameters. We centered our $\chi^2$ at $\Delta_{\text{NP}} = 0$ and $\Delta_{m_W}$ = 0 because we are assuming data to realize the SM expectation for the W-boson mass $\overline{m}_W$. We stress that we are testing the New Physics hypothesis with no prior on $\overline{m}_W$, as both $\Delta_{\text{NP}}$ and $m_{W}$ are floated.

On the contrary, the authors of \cite{Bandyopadhyay:2022bgx} {\it fixed} $m_W$ in the hypothesis to the EW fit prediction. The simultaneous fit to $m_W$ and NP that we perform here is thus a more general test of NP and has the added value to be independent of the EW fit results and the assumptions therein.

\begin{figure}
    \centering
\includegraphics[width=0.4\textwidth]{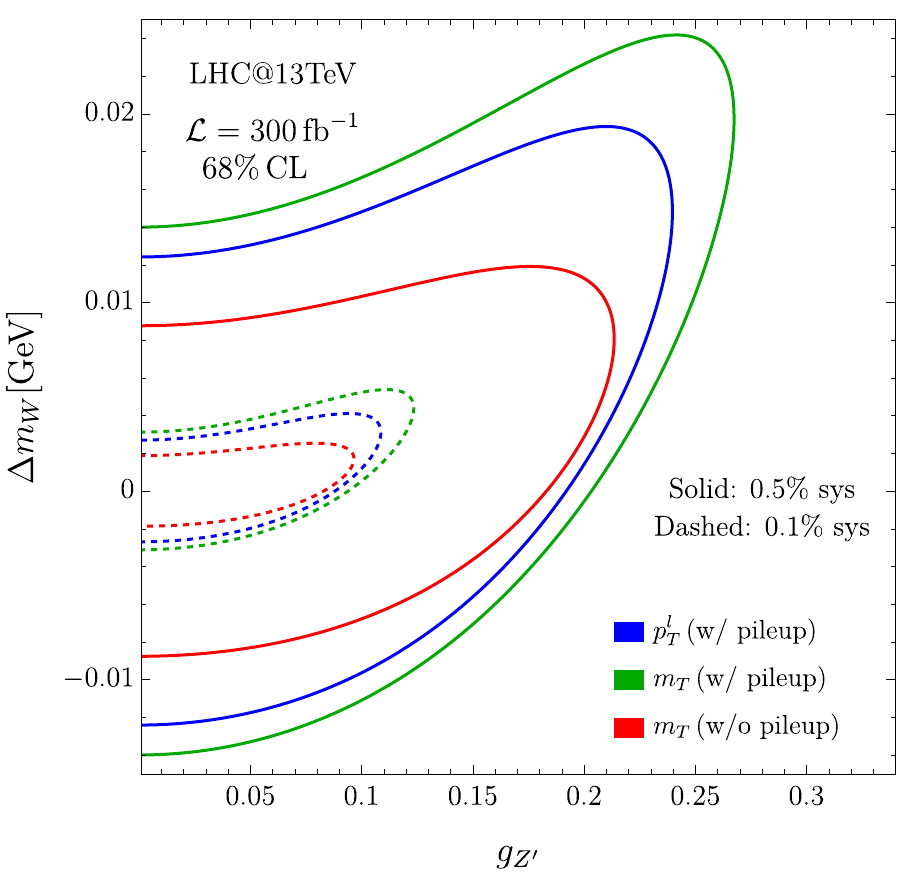}
\captionsetup{justification=Justified, singlelinecheck=off}
    \caption{$68\%$\ CL projected sensitivity to $L_{\mu}-L_{\tau}$ at LHC (ATLAS) ($m_{Z'} = 10\, \text{GeV}$).}
    \label{fig:mWvsGzLepto}
\end{figure}

The qualitatively new aspect of $\Delta_{m_W}$ being a floated parameter in Eq.~\eqref{Eq:ChiSqu} implies that with the same analysis we extract $m_W$ and test NP. The 2-dimensional fit in the $(\Delta_{m_W},\Delta_{\text{NP}})$ is reported in Fig.~\ref{fig:mWvsGzLepto} for $m_{Z'} = 10$ GeV. By assuming 0.5\% per-bin uncorrelated systematics and including the effect of pileup through $\textsc{Delphes}$, the ATLAS measured uncertainty is roughly reproduced.\footnote{The average number of pileup events per bunch crossing is $\langle \mu \rangle=50$.} Pileup has an impact on the $m_T$ distribution and on the resulting $m_W$ sensitivity. The $p_T^{\ell}$ distribution, on the contrary, is largely insensitive to pileup, hence we use it to draw more firm conclusions on features of our 2D-fit. 

The systematics on the kinematic distributions shown in \cite{ATLAS-CONF-2023-004} are below 0.5\%. Therefore, we also consider per-bin systematics of 0.1\%. The expected sensitivity to $m_W$ (at zero $g_{Z'}$) is slightly stronger than the current ATLAS 7~TeV $\mathcal{L}=4.6\text{ fb}^{-1}$ measurement~\cite{ATLAS-CONF-2023-004}. This is mainly because we are not including any source of correlated systematics, and we are assuming much larger statistics from a 13~TeV run with $\mathcal{L}=300 \text{ fb}^{-1}$. 

The distortion of the $p_T^\ell$ exclusion line (blue) at large values of $g_{Z'}$ implies a preference towards positive $\Delta_{m_W}$. This suggests that NP might in principle impact the sensitivity to $m_W$, possibly producing a shift in the extracted value  and/or affecting  the estimate of the associated uncertainty on $m_W$. Yet, the effect shown in Fig.~\ref{fig:mWvsGzLepto} is limited to only $\sim 10 \text{ MeV}$. However, a quantitative assessment of this effect requires the inclusion of the proper experimental setup and is beyond the scope of this \lett. The sensitivity to $g_{Z'}$ at $\Delta_{m_W}=0$ is only marginally affected by pileup, showing the robustness of the sensitivity to NP.

For completeness, we report in \cref{subfig:GzLeptoCDF2d} in the supplemental material an analogous study for CDF~\cite{CDF:2022hxs}. In this case, the effect of the NP in the $m_W$ determination is less pronounced, due to a sharper Jacobian peak related to the better control of the hadronic activity at CDF which anchors the $m_W$ fit more robustly.

We now turn to the test of the NP hypothesis. Assuming no prior knowledge on $\overline{m}_W$, the correct procedure to put bounds on NP is to marginalize on $\Delta m_W$ for each value of the NP parameters. This is shown in Fig.~\ref{subfig:zprime_money} for LHC ($\mathcal{L}=300 \text{ fb}^{-1}$) sensitivity projection. Prior knowledge on $\overline{m}_W$ (either from other measurements or from theory predictions) might impact the sensitivity to NP, as shown in Fig.~\ref{fig:mWvsGzLepto}. 

For this analysis, positively and negatively charged-muon events are added together, and $\chi^2$ for $p_T^\ell$ and $m_T$ are combined without correlation. Here, the sensitivity projections for CDF are also reported. The reach for $m_{Z'}\simeq 10$ GeV is competitive with the best probe for this model from a dedicated experiment (CCFR)~\cite{CCFR:1991lpl,Altmannshofer:2014pba}. Yet, it is remarkable that for a $10$ GeV $Z'$-boson, the $m_W$ measurement has the power to probe couplings $\sim few \times 0.01$, provided sufficient control of the systematics. Interestingly, less constrained models such as the ``neutrinophilic scalar'' of
\cite{de_Gouv_a_2020} or the ``Dirac neutrino portal'' \cite{Batell:2017cmf} fall in category (A). For the neutrinophilic scalar, we expect the $m_W$ measurement to be the best probe \cite{Agashe:toappear}.

\section{MSSM: Slepton-Sneutrino production}\label{Sec:SUSY}

We now turn to the minimal supersymmetric standard model (MSSM) \cite{Martin:1997ns}, which offers a simple irreducible ``background'' for the $m_W$ measurement: ``left-handed'' $SU(2)_L$ doublet slepton-sneutrino production, with subsequent decay into lepton plus only invisible particles (see Fig.~\ref{Fig:FeynDiag} right), 
\begin{align}
    p p \rightarrow \tilde{\ell}(\rightarrow \ell \, \tilde{\chi}^0_1) \, \tilde{\nu}_\ell \,.
    \label{Eq:MSSM}
\end{align}
In this scenario, both the sneutrino and neutralino are invisible, and either one could be the lightest stable particle (LSP).\footnote{When the lightest neutralino $\tilde{\chi}_1^0$ is the LSP, $\tilde{\ell} \rightarrow \ell\, \tilde{\chi}_1^0$, and $\tilde{\nu} \rightarrow \nu\, \tilde{\chi}_1^0$, as illustrated in Fig.~\ref{Fig:FeynDiag}, produces the $\ell+\met$ final state. If the sneutrino is the LSP (not shown), then $ \tilde{\chi}_1^0 \rightarrow \tilde{\nu}\, \nu$ also maintains the $\ell+\met$ final state.}
For simplicity, we assume that the other superpartners, including $SU(2)_L$ singlet -- or right-handed sleptons -- are heavy, thus having negligible cross-sections at the LHC. 

Sleptons lighter than $100$ GeV are excluded by LEP \cite{ALEPH:2001oot,ALEPH:2003acj,DELPHI:2003uqw,L3:2003fyi,OPAL:2003nhx}. Sleptons heavier than the LEP bound have negligible cross-section at the Tevatron so we do not consider CDF in this section. LHC searches for di-sleptons \cite{ATLAS:2022hbt,CMS:2023qhl} are sensitive to sleptons above the LEP bounds but suffer when the sleptons and $\tilde{\chi}^0$ are close by in mass. In particular, when the mass gap $m_{\tilde{\ell}}-m_{\tilde{\chi}_0}\sim m_W$, the lepton $p_T$ resembles that of the lepton from SM $W$-boson decay. This compressed region of parameter space is dominated by SM events and requires a dedicated analysis. In \cite{Curtin:2012nn,Curtin:2013gta} it has been proposed to use precision measurements to disentangle $WW$ events from di-slepton production. Yet, there is still some uncovered gap in the parameter space in the experimental results (see our summary of present constraints in \cref{subfig:susy_money}). Addressing this shortcoming of the present searches by filling this gap is a main result of this \lett.

The phenomenology of the process in eq.~\eqref{Eq:MSSM} belongs to category (C), since no on-shell $W$-boson is produced (see \cref{Fig:FeynDiag}). As shown in \cref{fig:Distrib}, NP produces a rather flat and extended $m_T$ distribution with a rising S/B ratio at ``high-$m_T$'', since the process is not initiated by the decay of a resonance.
The contamination in the $Z$-boson sample due to $p p \rightarrow \bar{\tilde{\ell}}\tilde{\ell} \rightarrow \bar{\ell} \ell \tilde{\chi}_0\tilde{\chi}_0$ is limited to $\mathcal{O}(10^{-5})$.

For this model, we follow the same procedure as in Sec.~\ref{Sec:lmultau} of marginalizing on $\Delta_{m_W}$ for varying NP parameters. For each point on the $m_{\tilde{\ell}}-m_{\tilde{\chi}_1^0}$ plane, $m_W$ is varied as an input in the template, and the minimum $\chi^2$ is obtained from the fit. 
The $m_W$ determination is largely governed by the peak positions of $p_T^\ell$ and $m_T$ spectra. Therefore, the rather flat kinematic distributions of NP contributions make a milder impact on the $m_W$ measurement than what is shown in \cref{fig:mWvsGzLepto}. 
Sensitivity projections are reported in \cref{subfig:susy_money} as functions of $(m_{\tilde{\ell}},m_{\tilde{\chi}_0})$. The sneutrino mass is fixed at the lowest allowed value in the MSSM, assuming the large $\tan \beta$ limit \cite{Martin:1997ns}.

Two sets of expected sensitivities are reported in \cref{subfig:susy_money}, corresponding to the inclusion or not of pileup. In both cases, the fitting range (see Tab.~\ref{tab:CUTS}) is chosen to cover part of the unexplored parameter space. Extending the range to ``high-$m_T$'', still keeping sufficient control of the systematics, might improve the sensitivity, as shown in Fig.~\ref{fig:Distrib}. However, far from the ``$m_W$'' region, systematics becomes more challenging. This is caused, for instance, by the limited $Z$-boson sample available for calibrations, or by the increasing backgrounds. 
The study of systematics outside of the range presently used for each kinematic distribution employed in the $m_W$ measurement can only be carried out by the experimental collaborations. Here we are pointing out the huge gain 
in sensitivity to NP that can be obtained by enlarging the fitting range. Ideally ATLAS and CMS experiments will find the best range of each kinematic variable for which the experiment can keep systematics under control so as to maximize the sensitivity to NP.

A major result of ours is that the same analysis used for the $m_W$ measurement, with only a slightly extended fitting range, can put new bounds and potentially discover new physics in an unexplored parameter space of MSSM. 

\begin{table}[]
    \centering
    \begin{tabular}{c|c|c|c}
    &ATLAS \cite{ATLAS:2017rzl,ATLAS-CONF-2023-004} ($\mu$) &$\tilde{\ell}_{\mu} \tilde{\nu}_{\mu}$&$L_{\mu}-L_{\tau}$ \\ \hline
    
        $p_T^{\ell}$ (GeV)  & $> 30$ (analysis) & $>30$&$>20$\\ 
          & $> 18$ (trigger) && \\

        $p_T^{\text{miss}}$&$>30$&$>30$&$>20$ \\
          
        $m_T$ (GeV)  & $> 60$  & $>60$ & $>40$\\ 
        
        $|\vec{u}_T|$ (GeV)& $<30$ & $<30$&$<30$\\
        
        $m_T$ range (GeV)  & $[60,100]$  & $[60,120]^*$&$[40,100]$ \\  &   & $[60,140]$&\\ 
        
        $p_T^{\ell}$ range (GeV)  & $[30,50]$  & $[30,60]^*$&$[20,50]$\\
          &   & $[30,70]$&
    \end{tabular}
    \captionsetup{justification=Justified, singlelinecheck=off}
    \caption{Kinematic range considered for our fit. $\vec{u}_T$ is  the hadronic recoil vector. The range with $*$ is considered when we include no pileup effects. We construct bins of $2$ GeV for $m_T$ and $1$ GeV for $p_T^\ell$ \cite{ATLAS-CONF-2023-004}. }
    \label{tab:CUTS}
\end{table}

\section{Conclusion}\label{Sec:Conclusion}

New physics resulting in $\ell+\met$ is an irreducible ``background'' for the $m_W$ measurement. The kinematic distributions arising from NP do not match those of the SM $W$-boson. Consequently, a simultaneous fit to NP parameters and $m_W$ is required to capture this contamination of NP. This more general procedure also tests the robustness of the extraction of $m_W$.

Concerning the sensitivity to NP, the inclusion of possible NP worsens the goodness of the fit of the data to (pure) SM template. This results in strong bounds on the NP hypothesis. Yet, given the underlying uncertainties, the distributions contaminated by NP can also modify the extracted value of $m_W$ (Fig.~\ref{fig:mWvsGzLepto}).

In this \lett, we followed this path through two examples: anomalous $W$-boson decay via an invisible $L_{\mu}-L_{\tau}$ $Z'$-boson and slepton-sneutrino production in the MSSM.
We find that the LHC, provided sufficient control of the systematics, is potentially sensitive to an uncovered parameter space of the MSSM and provides a competitive probe for the $L_{\mu}-L_{\tau}$ $Z'$-boson, as shown in Fig.~\ref{fig:money_plots}.
A faithful assessment of this effect requires precise simulations of the experimental environment.

The paradigm that we follow in this \lett~is general and applies to all NP scenarios producing $\ell + \met$, pinpointed in Sec.\ref{Sec:NPtoWMass}. This is postponed to a future publication \cite{Agashe:toappear}. 

\begin{acknowledgements}
The authors would like to thank Alberto Belloni,
Bodhitha Jayatilaka,
Rafael Lopes de Sa,
Sarah Eno,
Tao Han,
Philip Harris,
Jakub Kremer,
Patrick Meade,
Federico Meloni,
Javier Montejo Berlingen,
Pier Francesco Monni,
Felix Yu,
Gustavo Marques-Tavares for discussions. 
The work of K.~A., S.~A., L.~R.~and D.~S.~is supported by NSF Grant No.~PHY-2210361 and by the Maryland Center for Fundamental Physics. 
The work of D.~K. is supported by the DOE Grant No. DE-SC0010813. The work of A.~V.~K. is supported by the DOE Grant No. DE-SC0010007. The work of R.~F. is partially supported by Ministero dell'Universit\`a e della Ricerca MUR under the grant PRIN 202289JEW4.
\end{acknowledgements}
\newpage

\bibliography{paper}
\bibliographystyle{apsrev4-1}

\pagebreak
\widetext

\section*{Supplementary Material}
\subsection{The \texorpdfstring{$L_{\mu}-L_{\tau}$}{L\textmu - L\texttau} gauge boson at CDF}

We describe the CDF analysis for the $L_{\mu} - L_{\tau}$ $Z'$-boson model (Sec.~\ref{Sec:lmultau}). 
The cuts are slightly modified from those imposed by CDF for the measurement of $m_W$, shown in table~\ref{tab:CUTS_CDF}. 
The SM and the NP events, without detector effects, are simulated using \textsc{MadGraph5\_aMC@NLOv3.42} \cite{Alwall:2014hca} + \textsc{PYTHIA8.212} \cite{Bierlich:2022pfr}. 
We simulate detector effects such as smearing of lepton momentum and hadronic recoil, including pileup, following Ref.~\cite{CDF:2022hxs}, such that the distributions resemble those published by CDF. 

\begin{figure}
\begin{minipage}{0.48\textwidth}
    \centering
    \includegraphics[width=0.9\textwidth]{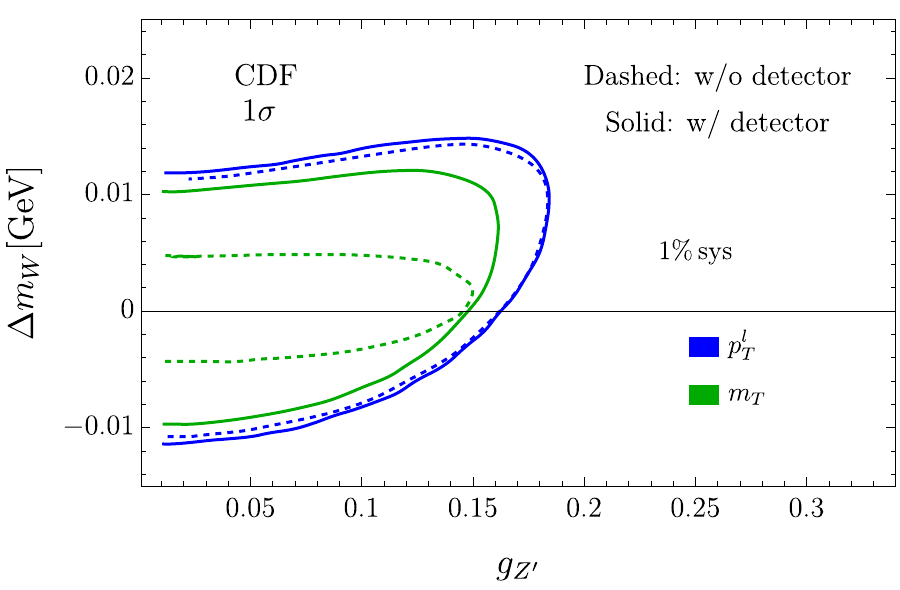}
    \subcaption{CDF 2D fit}
    \label{subfig:GzLeptoCDF2d}
\end{minipage}
\begin{minipage}{0.48\textwidth}
    \centering
    \includegraphics[width=0.9\textwidth]{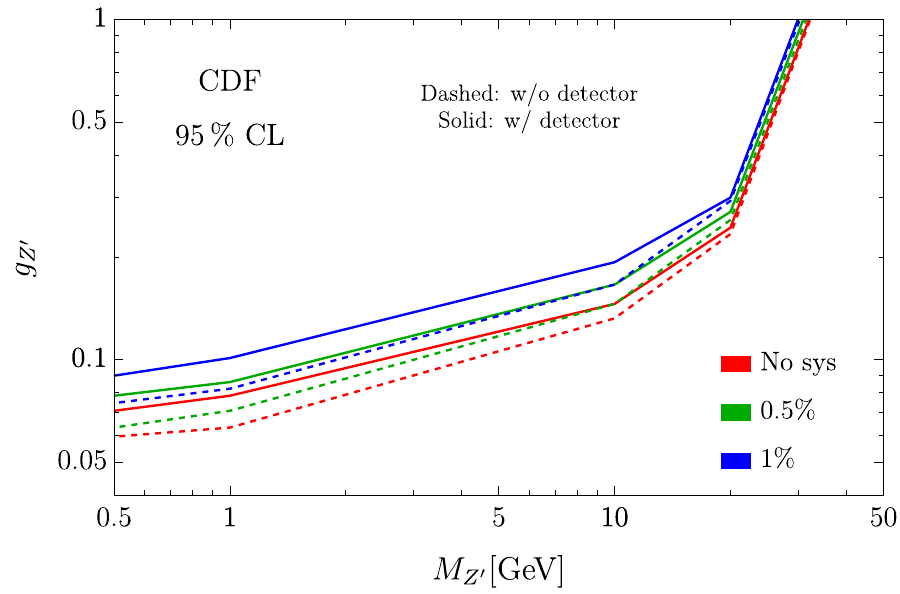}
    \subcaption{CDF projected sensitivity}
    \label{subfig:GzLeptoCDF}
\end{minipage}
    \caption{(a) 2D fit with 68\% CL projected sensitivity to $L_{\mu}-L_{\tau}$  for $m_{Z'}=10\text{ GeV}$ and (b) 95\% CL projected sensitivity to $L_{\mu}-L_{\tau}$ combining $p_T^\ell$, $p_T^{\rm miss}$, and $m_T$ at CDF. The solid lines and dashed lines represent the results when detector effects are included and when they are not included, respectively.}
\end{figure}

The simultaneous fit of $\Delta_{m_W}$ and $g_{Z'}$ is then performed for $m_T$, $p_T^\ell$ and $p_T^\text{miss}$ distributions. The $1\sigma$ contours for $m_T$ and $p_T^\ell$ are reported in Fig.~\ref{subfig:GzLeptoCDF2d} for $m_{Z'} = 10$ GeV. Notice the correlation between $g_{Z'}$ and the measured $m_W$ for both $m_T$ and $p_T$ contours in the presence of detector effects. When detector effects are not present, this correlation is absent for $m_T$. This suggests that the detector effects are responsible for larger $g_{Z'}$ preferring positive $\Delta_{m_W}$ for the fit using $m_T$. This correlation is small compared to ATLAS (in Fig.~\ref{fig:mWvsGzLepto}) because the pileup effects are much smaller at CDF. This holds true for $p_T^\ell$ as well since the Jacobian peak is sharper at CDF. These observations may change with a different treatment of the pileup in our simulation. 

Next, for CDF, projected sensitivity to the NP parameters are obtained by marginalizing over $\Delta_{m_W}$. The $\chi^2$ from $p_T^\ell$, $m_t$, and $p_T^{\rm miss}$ are combined (without correlation) to obtain exclusion contours.
Fig.~\ref{subfig:GzLeptoCDF} shows a sensitivity which is comparable to the ATLAS projections for this model. Since the NP signal populates the SM-dominated region, the reach at ATLAS is limited by systematics. Hence, even with lower statistics, CDF is competitive. 

\begin{table}[h!]
    \centering
    \begin{tabular}{c|c|c}
    &CDF ($\mu$) \cite{CDF:2022hxs}&$L_{\mu}-L_{\tau}$ \\ \hline
    
        $p_T^{\ell}$ (GeV)  & $[30, 55]$ (analysis) & $[20, 55]$\\ 
          & $> 18$ (trigger) &   \\

           $p_{T}^{\text{miss}}$& $[30, 55]$ & $[20, 55]$\\
        $m_T$ (GeV)  & $[60,100]$ & $[40,100]$\\ 
        
        $|\vec{u}_T|$ (GeV)& $<15$ & $<15$\\
        
        $m_T$ range (GeV)  & $[65,90]$  &$[40,90]$ \\ 
        
        $p_T^{\ell}$,$p_T^{\text{miss}}$ range (GeV)  & $[32,48]$  & $[20,48]$ 
    \end{tabular}
    \captionsetup{justification=Justified, singlelinecheck=off}
    \caption{Kinematic cuts and ranges considered for the fit. We construct bins of $0.5$ GeV for $m_T$ and $0.25$ GeV for $p_T^\ell$ and $p_T^{\text{miss}}$\cite{CDF:2022hxs}.}
    \label{tab:CUTS_CDF}
\end{table}

\end{document}